\documentclass[a4paper,11pt]{article}
\pdfoutput=1 %

\usepackage{jinstpub} %

\usepackage{lineno}
\notoc
\usepackage{multicol}
\usepackage{cprotect}

\newcommand{\wzurl}[1]{\href{#1}{\url{#1}}}

\usepackage[final,defaultcolor=red]{changes}

\title{\boldmath  Evolution of the data aggregation concepts for {STS} readout in the {CBM} Experiment}

\author[a]{Wojciech M. Zabołotny,\note{Corresponding author.}}
\author[b]{David Emschermann,}
\author[a]{Marek Gumiński,}
\author[a]{Michał Kruszewski,}
\author[b]{Jörg Lehnert,}
\author[a]{Piotr Miedzik,}
\author[b]{Walter F.J. Müller,}
\author[a]{Krzysztof Poźniak,}
\author[a]{Ryszard Romaniuk,}

\affiliation[a]{Institute of Electronic Systems, Warsaw University of Technology,\\
             Nowowiejska 15/19, 00-665 Warszawa, Poland}
\affiliation[b]{GSI - Helmholtzzentrum für Schwerionenforschung GmbH,\\
             Darmstadt, Germany}

\emailAdd{wojciech.zabolotny@pw.edu.pl}

\abstract{
The STS detector in the CBM experiment delivers data via multiple \replaced{e-links}{E-Links} connected to GBTX ASICs. In the process of data aggregation, that data must be received, combined into a smaller number of streams, and packed into so-called microslices containing data from specific periods. The aggregation must consider data randomization due to amplitude-dependent processing time in the FEE ASICs and different occupancy of individual \replaced{e-links}{E-Links}. During the development of the STS readout, the continued progress in the available technology affected the requirements for data aggregation, its architecture, and algorithms. The contribution presents considered solutions and discusses their properties.
}


\arxivnumber{2410.18733} %

\proceeding{TWEPP 2024\\
  29.09--4.10.2024}

\begin{document}
\maketitle
\flushbottom

\section{Introduction}
\begin{figure}[htbp]
\centering %
\includegraphics[width=.6\textwidth]{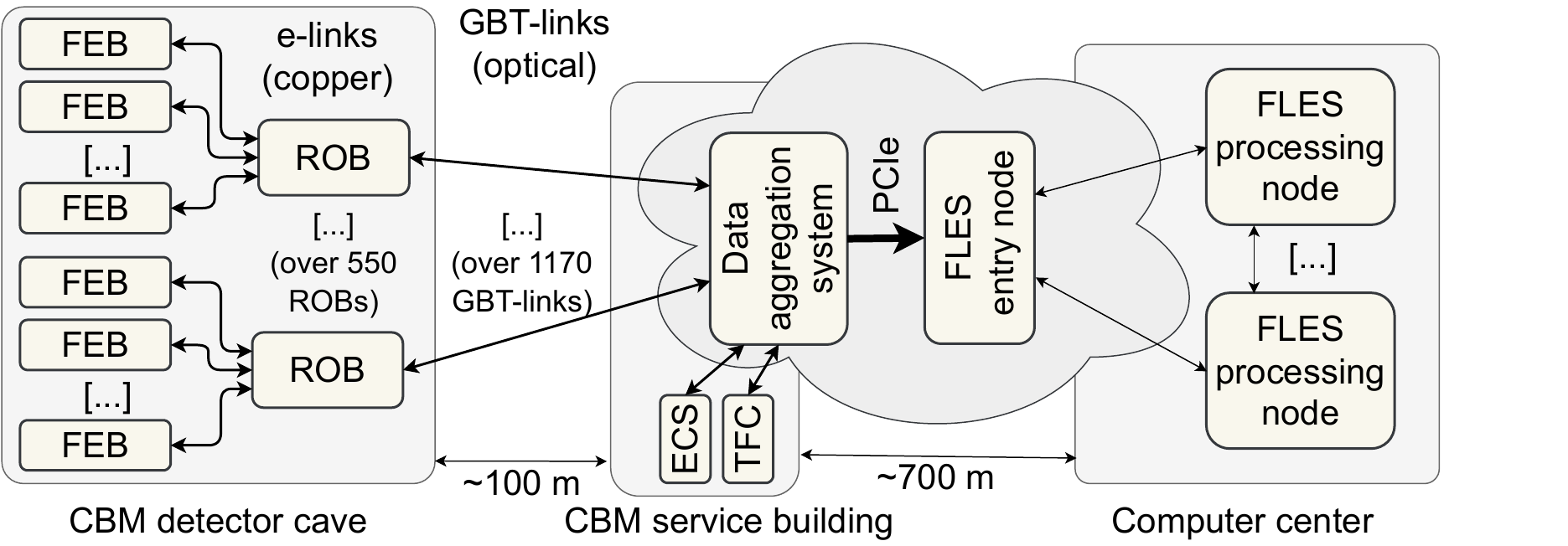}
\caption{\label{fig:gen-struc-daq}The general structure of the STS readout chain in the CBM experiment. The GBT-links transceivers must be located in the CBM service building. The location of the FLES entry node depends on the solution. (ECS - Experiment control system, TFC - Timing and Fast Control)}
\end{figure}

Silicon Tracking System (STS)~\cite{Heuser:54798} is one of the detectors in the CBM experiment prepared in FAIR/GSI in Darmstadt. The experiment uses triggerless free streaming data acquisition~\cite{cbm_collaboration_technical_2023}. The STS detector Front-End (FE) ASICs (SMX) mounted on Front-End Boards (FEBs) produce 24-bit data containing, among others, timestamped hits and epoch markers\footnote{
The SMX produces the 14-bit timestamp. Its resolution is 3.125 ns for a 320~Mb/s uplink rate, so the wrap-around period is 51.2~µs. The hit messages transmit only 10 lower bits of the timestamp. The epoch markers (TS-MSB) contain the bits 13--8, delivering the remaining 13--10 bits. Bits 9 and 8 are transmitted both in hits and epoch markers~\cite{kasinski_protocol_2016}. \added{The TS-MSB is protected with 4-bit CRC, and bits 13--8 are triplicated to enable recovery of a partially corrupted marker.} 
}, which are 8b/10b encoded~\cite{kasinski_protocol_2016} and sent at 320 Mb/s via more than 20000 e-links connected to GBTX ASICs~\cite{moreira_gbtx_2021} in readout boards (ROBs). 
The data are further transmitted via more than 1700 GBT-links at 4.8 Gb/s to the data aggregation system\footnote{
\replaced{Each e-link delivers maximally one 24-bit 8b/10b encoded data word every $30\textrm{~b}/320\textrm{~Mb/s}=93.75\textrm{~ns}$.
A single GBT-link transmits data from up to 14 e-links delivering up to $149.333\textrm{~Mwords/s}$.}
{Every single GBT-link may transmit the data from up to 14 e-links. Each e-link may deliver maximally one 24-bit 8b/10b encoded data word every $30\textrm{~b}/320\textrm{~Mb/s}=93.75\textrm{~ns}$. Therefore, we may expect up to $149.333\textrm{~Mwords/s}$ from each GBT-link.}
}.
\added{The STS will be placed in a confined space in a dipole magnet~\cite{Heuser:54798}. That significantly limits the available uplink bandwidth. Because STS is a tracking detector with significant redundancy, it is possible to maximize the bandwidth at the cost of accepting a small amount of corrupted data. Therefore, no CRC is used for uplink detector data in e-links, nor is forward error correction (FEC) used in the GBT uplinks.}\footnote{\added{The main reason for data corruption is the non-zero bit error rate in e-links. The hit data corruption is similar (but significantly less frequent) to the effect of noise or limited efficiency of the STS sensors (additional wrong hits, lost hits). However, data aggregation must consider the possible corruption of epoch messages, which is more serious as it affects the interpretation of a higher number of hits.
}}
\replaced{In the aggregation system, the}{Here, that} data must be received, combined into a smaller number of streams, and packed into so-called microslices containing data from specific time intervals~\cite[chapter 4.2.1]{cbm_collaboration_technical_2023}, which are later combined into time-slices used for event reconstruction~\cite{friese_event_2020}. The aggregation must consider that data are not ordered according to their timestamp due to readout delay caused by different occupancy of individual e-links and amplitude-dependent processing time in the FE ASICs~\cite{kasinski_back-end_2016,kasinski_characterization_2018}. Finally, the concentrated data must be delivered via the PCIe interface to the First Level Event Selector (FLES) entry node, connected via the InfiniBand network to FLES computing nodes in the Computer Center. The general structure of the STS readout chain is shown in Figure~\ref{fig:gen-struc-daq}.
During the development of the STS readout, the continued progress in the available technology affected the requirements for data aggregation, its architecture, and algorithms.
Therefore, multiple concepts have been implemented and verified.

\section{The first concept of the readout}
The first proposed STS readout version~\cite{lehnert_gbt_2017} assumed the use of the intermediate FPGA-based Data Processing Boards (DPB) in the MTCA.4 standard. 
The MTCA.4 crate provided the possibility to deliver high-speed TFC signals and an IPbus-based control interface. 
The available MTCA.4 interconnect infrastructure could be reused for non-local preprocessing based on data received by multiple boards.
The DPB output was connected via a 10 Gb/s Aurora link to the PCIe FLES interface boards (FLIB)~\cite{hutter_cbm_2017}.

\begin{figure}[htbp]
\centering %
\includegraphics[width=.6\textwidth]{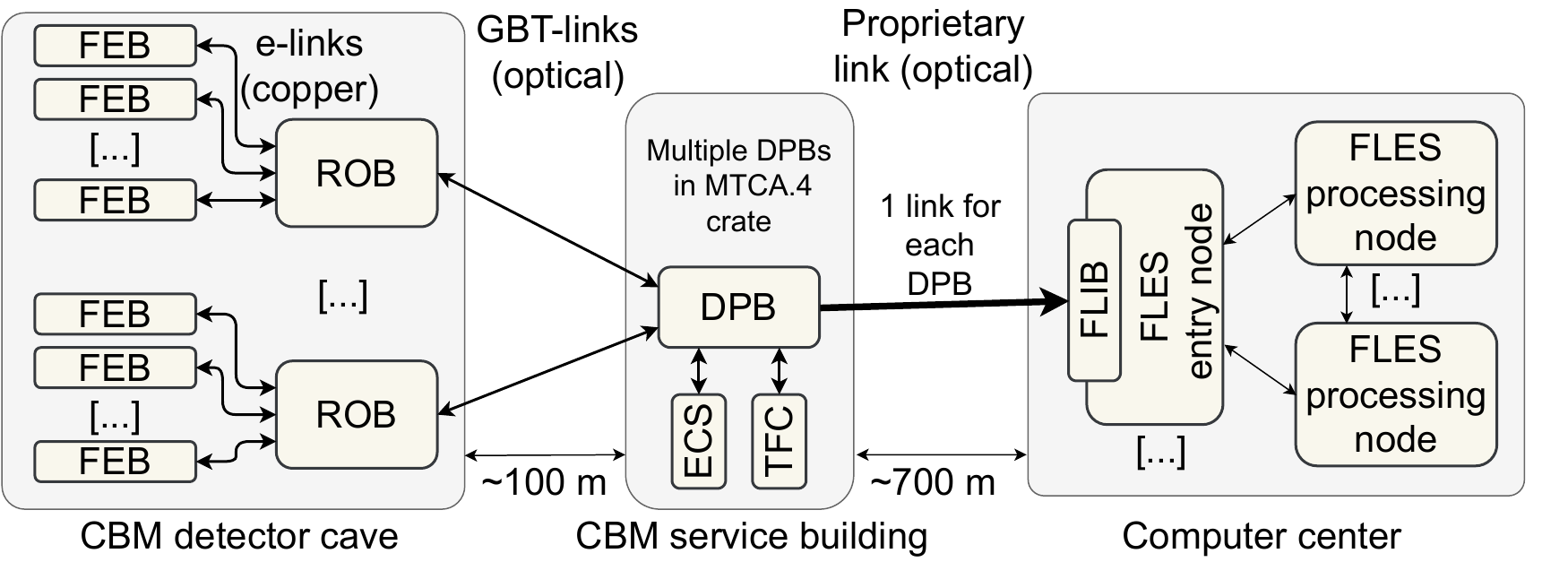}
\caption{\label{fig:dpb-based-readout}Block diagram of the first DPB-based prototype of the STS readout chain in the CBM experiment.}
\end{figure}

\begin{figure}[htbp]
\centering %
\includegraphics[width=.45\textwidth]{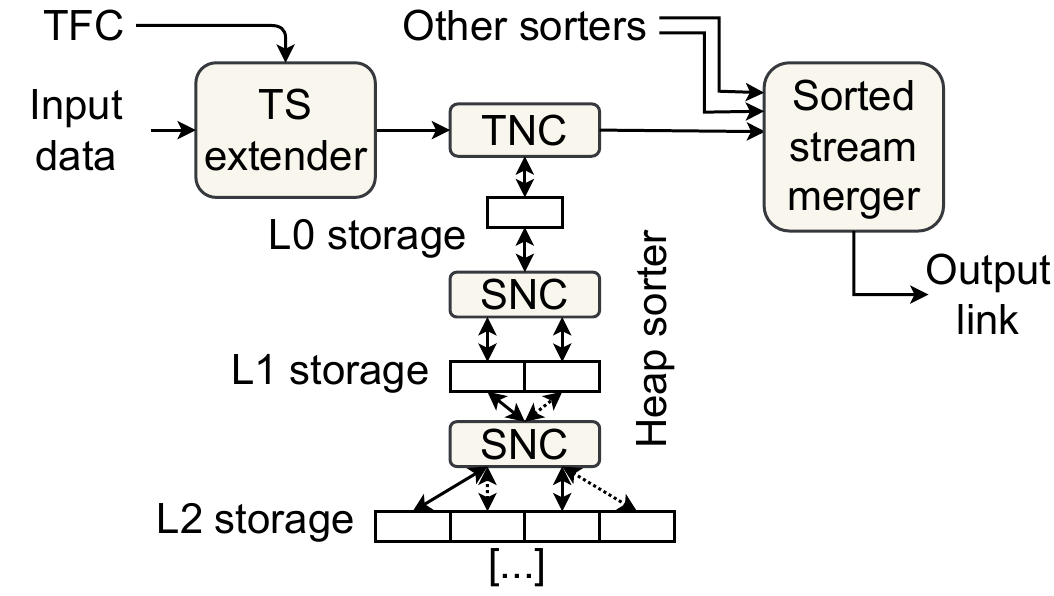}
\caption{\label{fig:heap-sorter}The data path with the heap sorter and stream merger.
(TS - timestamp, TNC - top node controller, SNC - sorting node controller). Figure based on~\cite{zabolotny_dual_2011}.}
\end{figure}

Each DPB board was expected to aggregate the data from 6 (possibly extensible to 8) GBT-links (see Figure~\ref{fig:dpb-based-readout}). For 24-bit data words, the expected maximum data bandwidth was 21.5 Gb/s for 6 GBT links (28.7 Gb/s for 8 GBT links) and was significantly higher than the available output bandwidth.
Therefore, a significant data preprocessing was considered to enable data volume reduction. At least, introducing
of a context-based data format was planned, where the number of bits needed to transfer a hit information could be reduced. 
Such processing required perfect sorting of the hit data according to their timestamp (TS). A dedicated TS extender\footnote{The TS extender restores additional bits of timestamps based on epoch markers received in the particular link.} and a heap sorter (see Figure~\ref{fig:heap-sorter}) blocks have been implemented for that purpose.
Unfortunately, in the beam tests, that solution appeared extremely sensitive to sorter overflow due to variations in the data rate caused by beam intensity fluctuations and the data timestamps corrupted by transmission errors in the e-links.
Also, implementing any FPGA-based non-local data processing appeared difficult due to high resource consumption.

\section{The second concept of the readout}
\begin{figure}[thbp]
\centering %
\includegraphics[width=.7\textwidth]{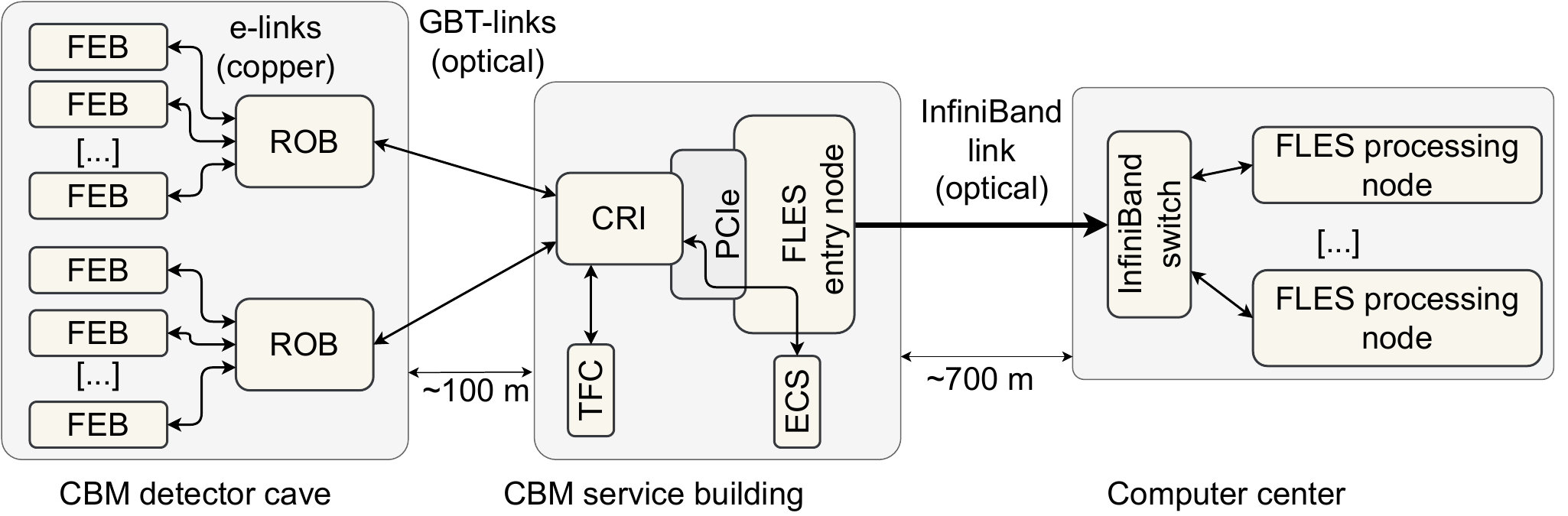}
\caption{\label{fig:cri-based-readout}Block diagram of the second prototype of the STS readout chain for the CBM experiment.}
\end{figure}

Avoiding the early non-local data processing enabled the elimination of the MTCA.4 crate and the intermediate FPGA layer. The functionalities of the DPB and FLIB boards have been integrated into new Common Readout Interface (CRI) boards~\cite[chapter 5.2]{cbm_collaboration_technical_2023}, which have been implemented as PCIe boards placed in the FLES entry nodes. That change required FLES entry nodes to be moved to the CBM service building, which resulted in significant advantages. The proprietary optical link with a rate limited by the transceivers in the FPGA in the DPB board could be replaced with a standard long-distance InfiniBand link to the FLES processing nodes in the Computer Center, which can run at higher speed and be easier maintained and upgraded (see Figure~\ref{fig:cri-based-readout}). The PCIe interface used to connect the CRI boards offers much higher bandwidth than the proprietary optical link\footnote{The theoretical maximum bandwidth of the PCIe Gen3x16 interface is 128~Gb/s. The theoretical maximum bandwidth of the PCIe Gen4x16 interface is 256~Gb/s. Even if it can't be fully utilized, it is much higher than the 10~Gb/s offered by the proprietary optical link offered by the DPB.}.

The CRI implements GBT links for ROB connectivity and the FLES Interface Module (FLIM) for PCIe. The hardware platform for the first CRI prototype was the FELIX BNL-712v2 board developed for the ATLAS experiment at CERN~\cite{paramonov_felix_2021} but used with CBM-developed firmware. The board may support a higher number of GBT-links - up to 47. Currently 24 GBT-Links are used for STS. The measured PCIe bandwidth ($2 \cdot 6.65~\textrm{GB/s} = 106.4~\textrm{Gb/s}$) is higher than the expected maximum input bandwidth ($86.02~\textrm{Gb/s}$ for 24 GBT-links), eliminating the need for context-based data aggregation, and thereby relaxing the requirement for perfect sorting.

\section{Bucket sorter-based data aggregation}
\begin{figure}[htbp]
\centering %
\includegraphics[width=.58\textwidth]{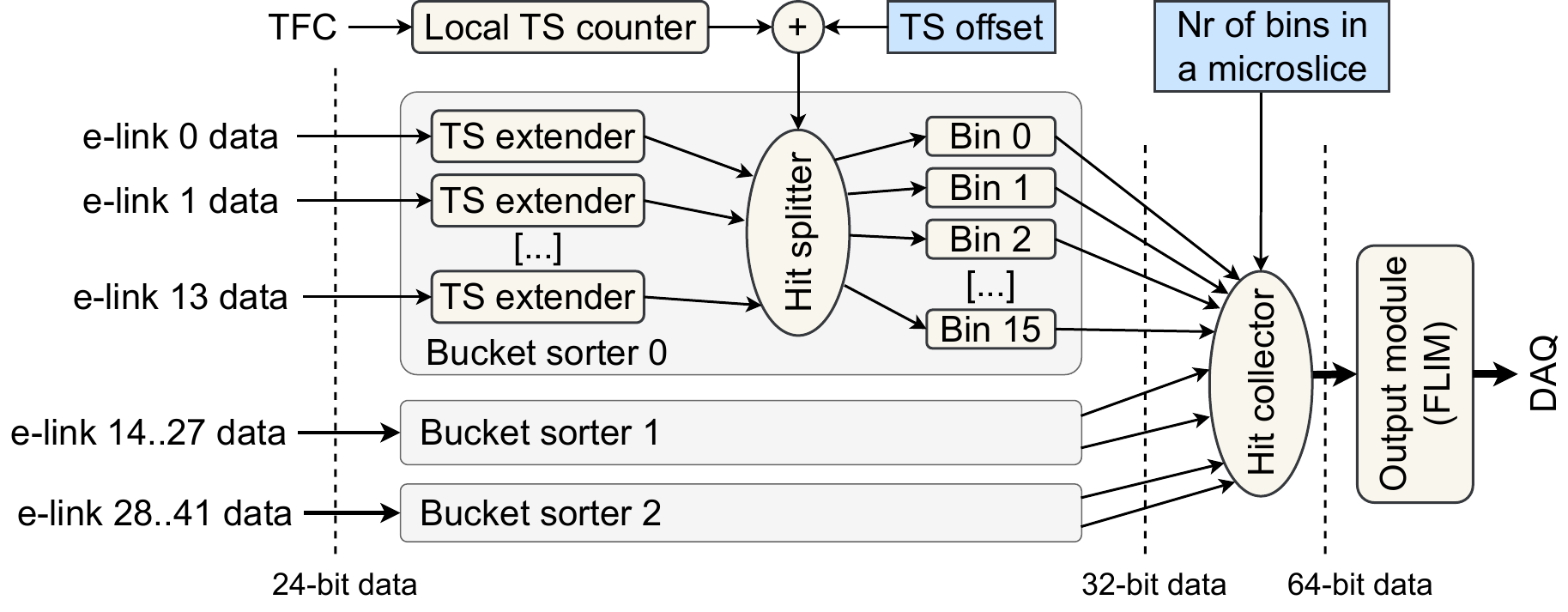}
\caption{\label{fig:bucket}Data aggregation with bucket sorters used in the second version of the CBM DAQ readout. Data from 3 GBT-links are delivered to 3 bucket sorters. The data are sorted into 16 bins based on 4 selected bits of their timestamp. The hit collector receives the data from the same bins from all 3 sorters and sends them to the output.
The 24-bit e-link data are supplemented with the source ID, forming 32-bit data. Finally, data are packed into 64-bit words used by FLIM. Figure based on~\cite[Figure B.5]{cbm_collaboration_technical_2023}.} 
\end{figure}

The relaxed requirements for data compression enabled the replacement of the heap sorter with the bucket sorter, providing partially sorted data~\cite[appendix B.1.4]{cbm_collaboration_technical_2023}. In that approach (see Figure~\ref{fig:bucket}), the data received from a group of 14 e-links are concentrated, and their TS is extended.  Then, four bits of TS are used to select the bin. The lower bits are ignored. The higher bits define the acceptable range of timestamps\footnote{
 The least significant bit of TS corresponds to 3.125~ns. Therefore, if, for example, bits 10 to 13 are chosen for bin selection, the ten lowest bits are ignored and the time period covered by each bin is $1024 \cdot 3.125\textrm{~ns} = 3.2~\textrm{~µs}$.
}.

This solution better handles intermittent peaks in the data rate. The data delivered to an already full bin are rejected, and the bin's ``data lost'' flag is set. The data with corrupted timestamps are rejected if their timestamp is outside the limit matching 16 available bins. Of course, the corruption of lower bits may lead to data being stored in an incorrect bin.
There is still a small risk of disturbing the operation of the TS extender by the corrupted epoch markers in the data stream.
The main problem with that solution is the static allocation of the same memory amount for each bin. In case of data rate fluctuations, it is not possible to compensate for the higher memory occupancy in one bin with the lower memory usage in another bin.
In the beam tests, with the bucket sorter handling data from a single GBT-link for reasonable bin duration (3.2µs) and  memory size (1024 words), the data loss due to bin overflow occurred unacceptably often.

\section{Aggregation of data with simple concentration}
The previously described solutions heavily depend on the timestamps contained in received data, which makes them sensitive to data corruption. Additionally, they significantly modify the data stream. In case of problems, reconstructing original data and diagnosing the problem is impossible. The debugging requires adding a dedicated diagnostic mode (which increases the FPGA resource consumption) or providing a special diagnostic firmware (which requires reconfiguring the FPGAs). Therefore, yet another aggregation concept was tested.
The last progress in PCIe technology (Gen4 and Gen5) further increases the bandwidth available for data transmission. Based on that, a new concept utilizing the simple concentration of data has been created.
Other experiments have also used the concepts of using FPGA-based boards as almost transparent data concentrators. For the LHCb experiment at CERN, the PCIe40 board~\cite{cachemiche_pcie-based_2016} was developed. The transparent data concentration has also been proposed for the ATLAS experiment at CERN~\cite{anderson_new_2016}, and the boards developed for that project are used as the hardware platform for prototype CRI boards in the CBM readout.

In the simple concentration approach, the data words received from a group of e-links (up to 15) are serialized at 160 MHz, supplemented with the source ID (number of e-link and number of the GBT link), creating the 32-bit words. However, the PCIe output module uses a wider data word (256, 512, or 1024 bits). Therefore, the data from multiple groups may be stored in a single output word. A dedicated high-speed concentrator~\cite{zabolotny_scalable_2023} has been created to pack such data into wider words without leaving empty places and wasting clock cycles. 

\begin{figure}[htbp]
\centering %
\includegraphics[width=.8\textwidth]{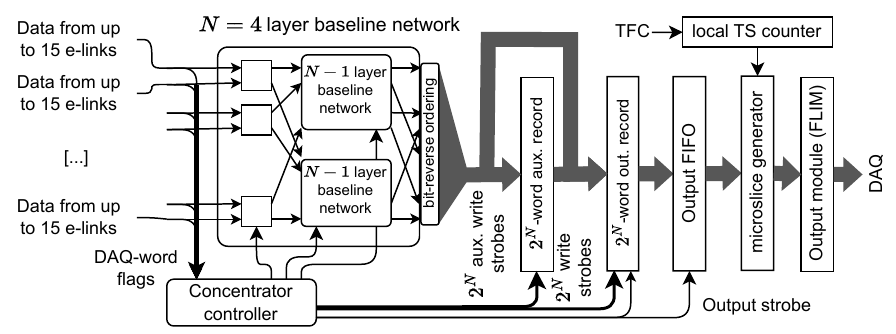}
\caption{\label{fig:concentrator} System performing the simple concentration of the data in STS readout. Figure based on~\cite{zabolotny_scalable_2023}. As proven in~\cite{zabolotny_scalable_2023}, the N-layer baseline network is capable of reordering the data from up to $2^N$ inputs so that the DAQ words from consecutive inputs are routed to consecutive (modulo 
$2^N$, and in bit-reversed order) outputs. Non-DAQ words are skipped. That enables packing the DAQ words into the output record without leaving empty spaces,  preserving their order according to their arrival time at the concentrator and the input number. In the particular clock period, the output record may be partially occupied. In that case, some data are preserved in the auxiliary record until the next clock period.}
\end{figure}

In that approach (see Figure~\ref{fig:concentrator}), the boundaries of the microslices are determined by the arrival time of the data, hence completely eliminating the influence of corrupted data\footnote{
Simplified generation of microslices may lead to a situation where data from the same period are stored in neighboring microslices. However, later analysis uses timeslices built as overlapping sequences of microslices~\cite[chapter 4.4]{cbm_collaboration_technical_2023}. Therefore, each timeslice contains all the data from a given time interval despite potential time disorder in microslices.
}. 
The original data stream may be fully reconstructed, so the detection of anomalies \added{(including those related to data corrupted in readout)} may be implemented in the software.
One modification of the original data stream is necessary if an individual e-link does not deliver data for a prolonged time. In such a case, artificially created epoch markers must be inserted.

That solution has been successfully tested with a DMA engine developed for the GERI board~\cite{electronics12040883} with a 256-bit output word, and in a simplified form (with a 64-bit output word) in the first prototype of the CRI board.
The tests have confirmed that this aggregation scheme offers the best handling of high data rates among the described solutions.

\section{Conclusions}
The preparation of the CBM experiment inspired the development and testing of various methods for the aggregation of detector data. The selection of the particular method depended on the currently available technology. The currently selected solution utilizes the progress in the FPGA and PCIe technology to ensure almost transparent transmission of the detector-produced data stream, eliminating the need for a separate diagnostic mode. All information contained in the original data is available for software processing in the FLES computing nodes. However, the effort invested in developing earlier solutions is not void. The elaborated solutions may be reused in other systems where perfect or partial data sorting in the FPGA layer is necessary. Most solutions described in the paper are open-source and available for the HEP community.

\acknowledgments

The work has been partially supported by GSI and ISE. Part of the work was done in the project that received funding from the European Union’s Horizon 2020 research and innovation programme under grant agreement no 871072, and from Polish Ministry of Education and Science programme ``Premia na Horyzoncie 2''.

\bibliography{evolution}

\providecommand{\href}[2]{#2}\begingroup\raggedright\begin{thebibliography}{10}

\bibitem{Heuser:54798}
J.~Heuser, W.~Müller, V.~Pugatch, P.~Senger, C.J.~Schmidt, C.~Sturm et~al.,
  eds., \emph{[{GSI} {R}eport 2013-4] {T}echnical {D}esign {R}eport for the
  {CBM} {S}ilicon {T}racking {S}ystem ({STS})}, GSI, Darmstadt (2013).

\bibitem{cbm_collaboration_technical_2023}
{CBM Collaboration}, \emph{Technical {Design} {Report} for the {CBM} {Online}
  {Systems} – {Part} {I}, {DAQ} and {FLES} {Entry} {Stage}}, GSI
  Helmholtzzentrum fuer Schwerionenforschung, GSI, Darmstadt (2023),
  \href{https://doi.org/10.15120/GSI-2023-00739}{10.15120/GSI-2023-00739}.

\bibitem{kasinski_protocol_2016}
K.~Kasinski, R.~Szczygiel, W.~Zabolotny, J.~Lehnert, C.~Schmidt and W.~Müller,
  \emph{A protocol for hit and control synchronous transfer for the front-end
  electronics at the {CBM} experiment},
  \href{https://doi.org/10.1016/j.nima.2016.08.005}{\emph{Nuclear Instruments
  and Methods in Physics Research Section A: Accelerators, Spectrometers,
  Detectors and Associated Equipment} {\bfseries 835} (2016) 66}.

\bibitem{moreira_gbtx_2021}
P.~Moreira, J.~Christiansen, K.~Wyllie, P.~Moreira, S.~Baron, S.~Bonacini
  et~al., ``{GBTX} manual; {V0}.18 draft.''
  \wzurl{https://cds.cern.ch/record/2809057}, 2021.

\bibitem{friese_event_2020}
V.~Friese, \emph{Event {Reconstruction} in the {Tracking} {System} of the {CBM}
  {Experiment}}, \href{https://doi.org/10.1051/epjconf/202022601004}{\emph{EPJ
  Web of Conferences} {\bfseries 226} (2020) 01004}.

\bibitem{kasinski_back-end_2016}
K.~Kasinski et~al., \emph{Back-end and interface implementation of the
  {STS}-{XYTER2} prototype {ASIC} for the {CBM} experiment},
  \href{https://doi.org/10.1088/1748-0221/11/11/C11018}{\emph{Journal of
  Instrumentation} {\bfseries 11} (2016) C11018}.

\bibitem{kasinski_characterization_2018}
K.~Kasinski, A.~Rodriguez-Rodriguez, J.~Lehnert, W.~Zubrzycka, R.~Szczygiel,
  P.~Otfinowski et~al., \emph{Characterization of the {STS}/{MUCH}-{XYTER2}, a
  128-channel time and amplitude measurement {IC} for gas and silicon
  microstrip sensors},
  \href{https://doi.org/10.1016/j.nima.2018.08.076}{\emph{Nuclear Instruments
  and Methods in Physics Research Section A: Accelerators, Spectrometers,
  Detectors and Associated Equipment} {\bfseries 908} (2018) 225}.

\bibitem{lehnert_gbt_2017}
J.~Lehnert et~al., \emph{{GBT} based readout in the {CBM} experiment},
  \href{https://doi.org/10.1088/1748-0221/12/02/C02061}{\emph{Journal of
  Instrumentation} {\bfseries 12} (2017) C02061}.

\bibitem{hutter_cbm_2017}
D.~Hutter, J.d.~Cuveland and V.~Lindenstruth, \emph{{CBM} {First}-level {Event}
  {Selector} {Input} {Interface} {Demonstrator}},
  \href{https://doi.org/10.1088/1742-6596/898/3/032047}{\emph{Journal of
  Physics: Conference Series} {\bfseries 898} (2017) 032047}.

\bibitem{zabolotny_dual_2011}
W.M.~Zabołotny, \emph{Dual port memory based {Heapsort} implementation for
  {FPGA}},  in \emph{Proc. {SPIE}}, R.S.~Romaniuk, ed., vol.~8008, (Wilga,
  Poland), pp.~80080E--80080E--9, June, 2011,
  \href{https://doi.org/10.1117/12.905281}{DOI}.

\bibitem{paramonov_felix_2021}
A.~Paramonov, \emph{{FELIX}: the {Detector} {Interface} for the {ATLAS}
  {Experiment} at {CERN}},
  \href{https://doi.org/10.1051/epjconf/202125104006}{\emph{EPJ Web of
  Conferences} {\bfseries 251} (2021) 04006}.

\bibitem{cachemiche_pcie-based_2016}
J.~Cachemiche, P.~Duval, F.~Hachon, R.L.~Gac and F.~Réthoré, \emph{The
  {PCIe}-based readout system for the {LHCb} experiment},
  \href{https://doi.org/10.1088/1748-0221/11/02/P02013}{\emph{Journal of
  Instrumentation} {\bfseries 11} (2016) P02013}.

\bibitem{anderson_new_2016}
J.~Anderson, A.~Borga, H.~Boterenbrood, H.~Chen, K.~Chen, G.~Drake et~al.,
  \emph{A new approach to front-end electronics interfacing in the {ATLAS}
  experiment},
  \href{https://doi.org/10.1088/1748-0221/11/01/C01055}{\emph{Journal of
  Instrumentation} {\bfseries 11} (2016) C01055}.

\bibitem{zabolotny_scalable_2023}
W.M.~Zabołotny, \emph{Scalable {Data} {Concentrator} with {Baseline}
  {Interconnection} {Network} for {Triggerless} {Data} {Acquisition}
  {Systems}},
  \href{https://doi.org/10.3390/electronics13010081}{\emph{Electronics}
  {\bfseries 13} (2023) 81}.

\bibitem{electronics12040883}
W.M.~Zabołotny, \emph{Versatile {DMA} engine for high-energy physics data
  acquisition implemented with {High-Level Synthesis}},
  \href{https://doi.org/10.3390/electronics12040883}{\emph{Electronics}
  {\bfseries 12} (2023) }.

\end{thebibliography}\endgroup
\bibliographystyle{JHEP}

\end{document}